\documentclass{article}

\usepackage{times}
\usepackage{helvet}
\usepackage{courier}
\usepackage{epsfig}
\usepackage[TABBOTCAP]{subfigure}
\usepackage{tabularx}
\usepackage{graphicx}
\usepackage{color}
\usepackage{xspace}
\usepackage{thumbpdf}
\usepackage{listings}
\usepackage{verbatim}
\usepackage{booktabs}
\usepackage{colortbl}
\usepackage{amsmath}
\usepackage{amssymb}
\usepackage{mathtools}
\usepackage{multirow}
\usepackage{authblk}

\newcommand{\one}{({\em  i}\/)\xspace}
\newcommand{\two}{({\em  ii}\/)\xspace}

\newcommand{\eg}{{\em e.g.}\xspace}
\newcommand{\etc}{{\em etc}\xspace}
\newcommand{\ie}{{\em i.e.}\xspace}

\newcommand{\vs}{{\em vs.}\xspace}
\newcommand{\cf}{{\em cf.}\xspace}

\newcommand{\stweeler}{{\textbf{\textit{Stweeler}}}\xspace}

\newcommand{\gtenm}{{${\bf G}_{\bf 10M+}$}\xspace}
\newcommand{\gonem}{{${\bf G}_{\bf 1M}$}\xspace}
\newcommand{\ghundredk}{{${\bf G}_{\bf 100k}$}\xspace}
\newcommand{\gonek}{{${\bf G}_{\bf 1k}$}\xspace}

\newcounter{ZANumberOfComments}
\stepcounter{ZANumberOfComments}

\newcommand{\zafar}[1]{\textcolor{red}{\small \bf [ZG\#\arabic{ZANumberOfComments}\stepcounter{ZANumberOfComments}: #1]}}

\newcounter{GTNumberOfComments}
\stepcounter{GTNumberOfComments}

\newcommand{\gareth}[1]{\textcolor{blue}{\small \bf [GT\#\arabic{GTNumberOfComments}\stepcounter{GTNumberOfComments}: #1]}}

\newcounter{RZNumberOfComments}
\stepcounter{RZNumberOfComments}

\newcommand{\reza}[1]{\textcolor{cyan}{\small \bf [RF\#\arabic{RZNumberOfComments}\stepcounter{RZNumberOfComments}: #1]}}

\newcounter{LWNumberOfComments}
\stepcounter{LWNumberOfComments}

\newcommand{\liang}[1]{\textcolor{magenta}{\small \bf [LW\#\arabic{LWNumberOfComments}\stepcounter{LWNumberOfComments}: #1]}}

\newif\ifcommentoff

\ifcommentoff
	\renewcommand{\zafar}[1]{}
	\renewcommand{\gareth}[1]{}
	\renewcommand{\reza}[1]{}
	\renewcommand{\liang}[1]{}
\fi

\providecommand{\keywords}[1]{\textbf{\textit{Index terms---}} #1}

\begin{document}

\title{An in-depth characterisation of Bots and Humans on Twitter\\{\em \normalsize A Technical Report}}

\author[1]{Zafar Gilani \thanks{Email: \texttt{szuhg2@cam.ac.uk}; Corresponding author}}
\author[2]{Reza Farahbakhsh}
\author[3]{Gareth Tyson}
\author[1]{Liang Wang}
\author[1]{Jon Crowcroft}
\affil[1]{Computer Laboratory, University of Cambridge}
\affil[2]{Institut Mines Telecom Paris}
\affil[3]{Queen Mary University of London}

\date{}


\maketitle

\begin{abstract}
Recent research has shown a substantial active presence of bots in online social networks (OSNs).
In this paper we utilise our past work on studying bots (\stweeler) to comparatively analyse the usage and impact of bots and humans on Twitter, one of the largest OSNs in the world.
We collect a large-scale Twitter dataset and define various metrics based on tweet metadata.
We divide and filter the dataset in four popularity groups in terms of number of followers.
Using a human annotation task we assign `bot' and `human' ground-truth labels to the dataset, and compare the annotations against an online bot detection tool for evaluation.
We then ask a series of questions to discern important behavioural bot and human characteristics using metrics within and among four popularity groups.
From the comparative analysis we draw important differences as well as surprising similarities between the two entities, thus paving the way for reliable classification of automated political infiltration, advertisement campaigns, and general bot detection.
\end{abstract}


\section{Introduction}
\label{sec:introduction}

{\em Bots} (automated agents) exist in vast quantities in online social networks (OSNs).
They are created for a number of purposes, such as news, emergency communication, marketing, link farming,\footnote{Link farming -- \scriptsize\texttt{http://bit.ly/2cXhfBv}} political infiltration,\footnote{Bots distort U.S. presidential election -- \scriptsize\texttt{http://bit.ly/2l3VzGf}} spamming and spreading malicious content.
According to independent research, 51.8\% of all Web traffic is generated by bots.\footnote{Bot traffic report 2016 -- \scriptsize\texttt{http://bit.ly/2kzZ6Nn}}
Similarly, OSNs have seen a massive surge in their bot population; Twitter reported\footnote{Twitter's 2014 Q2 SEC -- \scriptsize\texttt{http://bit.ly/1kBx4M8}} in 2014 that 13.5 million accounts (5\% of the total user population back then) were either fake or spam.
The rise of bots on Twitter is further evidenced by a number of studies~\cite{lee2011seven,Wu2013,Edwards2014372,CresciPPST17}, and articles discussing bots.\footnote{Bots in press and blogs -- \scriptsize\texttt{http://bit.ly/2dBAIbB}}


This constitutes a radial shift in the nature of content production, which has traditionally been the realm of human creativity (or at least intervention).
Although there have been past studies on bots (\S\ref{sec:relatedwork}), we are particularly interested in exploring their role in the wider social ecosystem, and how their behavioural characteristics differ from humans.
This is driven by many factors.
The limited cognitive ability of bots clearly plays a major role, however, it is also driven by their diverse range of purposes, ranging from curating news to answering customer queries.
This raises a number of interesting questions regarding how these bots operate, interact and affect online content production:
What are the typical behaviours of humans and bots, in terms of their own activities as well as the reactions of others to them?
What interactions between humans and bots occur?
How do bots affect the overall social activities?
The understanding of these questions can have deep implications in many fields such as social media analysis, systems engineering, \etc.

To answer these questions, we have performed a large-scale measurement and analysis campaign on Twitter (\S\ref{sec:methodology}).\footnote{Datasets from this study will be made available upon notification.}
We focus on bots in Twitter because it largely exposes public content, and past studies indicate a substantial presence of bots~\cite{Chu2010}.
Addressing existing problems with automated bot detection algorithms, we utilise a human annotation task to manually identify bots, providing us with a large ground-truth for statistical analyses.
We analyse the most descriptive features from the dataset, as outlined in a social capitalist study \cite{dugue2015reliable}, including six which have not been used in the past to study bots.
Through our comprehensive approach, we offer a new and fundamental understanding of the characteristics of bots \vs humans, observing a number of clear differences (\S\ref{sec:mannersmakethbot}).
For example, we find that humans generate far more novel content, while bots rely more on retweeting.
We also observe less intuitive trends, such as the propensity of bots to tweet more URLs, and upload bulkier media (\eg images).
We also see divergent trends between different popularity groups (based on follower counts), with, for example, popular celebrities utilising bot-like tools to manage their fanbase.

We analyse the social interconnectedness of bots and humans to characterise how they influence the wider Twittersphere.
We observe that, although human contributions are generally considered more important via typical metrics (\eg number of likes, retweets), bots still sustain significant influence.
Our experiments confirm that the removal of bots from Twitter could have serious ramifications for information dissemination and content production on the social network.
Specifically, we simulate content dissemination to find that bots are involved in 54.59\% of all information flows (defined as the transfer of information from one user to another user).
As well as providing a powerful underpinning for future bot detection methods, our work makes contributions to the wider field of social content automation.
Such understanding is critical for future studies of social media, which are often skewed by the presence of bots.

\section{Related Work}
\label{sec:relatedwork}

Two main streams of research are relevant to our work: \one social, demographical and behavioural analyses of either bots or humans; and \two the impact of bots in limited social environments.
Bot detection per se is not the focus of this paper.
Many such techniques are focussed on discerning anomalous from normal, spam from non-spam, and fake from original, but they fail to distinguish (or compare) the types of users.
Note that a user can be a human and still be a spammer, and an account can be operated by a bot and still be benign.

\textbf{Social analysis of bots or humans.}
Most related to our work are behavioural studies of bots or humans.
For example, \cite{freitas2015reverse} studied the \emph{infiltration strategies} of social bots on Twitter using a manual approach.
They use three metrics to quantify the infiltration of social bots: followers, popularity score, and message-based interaction (other users favouriting, retweeting, replying or mentioning the bot).
They found that bots can successfully evade Twitter defences (only 38 out of their 120 bots got suspended over the course of 30 days).
The authors also found that bots can successfully infiltrate Twitter: 20\% of the bots had more than a 100 followers.
A similar work \cite{Boshmaf:2011:SNB:2076732.2076746} studied users' behaviours in response to bot infiltration, calculating an infiltration success rate of 80\%.
They concluded that a successful infiltration can result in privacy breaches (to expose more than publicly available data), and that the security defences of Facebook are ineffective in detecting infiltration.

Unlike these works, we do not aim to monitor the success of bot infiltration.
Rather we are interested in understanding the behavioural differences of bots and humans.
That said, there is work that has inspected bot or human behaviour in isolation.
For example, \cite{5428313} examined the retweet behaviour of people, focussing on {\em how people tweet}, as well as {\em why and what people retweet}.
The authors found that participants retweet using different styles, and for diverse reasons (\eg for others or for social action).
This is relevant to our own work, as we also study retweets.
In contrast, we directly compare retweet patterns of bots and humans (rather than just humans).

Thus, our work provides further insights on important differences and striking similarities between bots and humans in terms of {\em account lifetime}, {\em content creation}, {\em content popularity}, {\em entity interaction}, {\em content consumption}, {\em account reciprocity}, and {\em content dissemination}.
To the best of our knowledge, we are the first to perform this methodical comparison of key metrics across these two types of Twitter accounts.


\textbf{Social influence of bots.}
The above studies primarily inspect the characteristics of bots.
There has also been work inspecting the social influence of bots, \ie how other users react to them.
In \cite{aiello2014people}, the authors use a bot on aNobii, a social networking site aimed at readers, to explore the {\em trust}, {\em popularity} and {\em influence} of bots.
They show that gaining popularity does not require individualistic user features or actions, but rather simple social probing (\ie bots following and sending messages to users randomly).
The authors also found that an account can circumvent trust if it is popular (since popularity translates into influence).
In some cases their bot on aNobii was widely mistaken as a human and its activity triggered creation of groups and opinion polarisation, thus raising concerns about privacy violations and evoking fear of being controlled.
The results, again, confirm that bots can have a profound effect on online social media environments.
Closely related is~\cite{wagner2012social}, which develops models to identify users who are {\em susceptible} to social bots, \ie likely to follow and interact with bots.
The authors use a dataset from the Social Bot Challenge 2011, and make a number of interesting findings, \eg that users who employ more negation words have a higher susceptibility level.
We take inspiration from this work and extend exploration to the Twitter platform.
However, instead of infiltrating a social network with ``honeypot'' bot(s), we study the characteristics of existing bots.
We argue that this provides far broader vantage into real bot activities.
Hence, unlike studies that focus on the influence of individual bots (\eg the Syrian Civil War~\cite{abokhodair2015dissecting}), we gain perspective on the wider spectrum of how bots and humans operate, and interact.
Hence, we are not looking at how bots influence individual topics~\cite{Savage:Botivist}.

\section{Methodology}
\label{sec:methodology}
We use and build upon our past work \stweeler\footnote{\stweeler -- \scriptsize\texttt{https://github.com/zafargilani/stcs}} \cite{Gilani:2016:SFT:2872518.2889360, Gilani:2017} for data collection, pre-processing, human annotation, and analysis.
We define a `bot' as any account that {\em consistently} involves automation over the observed period, \eg use of the Twitter API or other third party tools, performing actions such as automated likes, tweets, retweets, \etc.
Note that a {\em tweet} is an original status and not a retweet, a {\em retweet} is a tweet which has `RT' in text, and a {\em status} is either a tweet or a retweet.
Also note that {\em content} on Twitter is limited to whatever is contained within a tweet: text, URL, image, and video.
We will explore this further under content generation (\S \ref{subsec:contentcreation}), content popularity (\S \ref{subsec:contentpopularity}), and content consumption (\S \ref{subsec:contentconsumption}).

\subsection{Data Collection}
\label{subsec:datacollection}

We selected Twitter because it is open, large-scale and is known to contain a wide breath of bot activities.
We collect data on bot and human behaviour for 30 days in April 2016 from the Streaming API.
Note that every single action is recorded as a {\em tweet} (status) on Twitter, whether a tweet, retweet, reply or mention.
Since we collect all the tweets within a time period $T$, we are certain we have comprehensive insights within $T$.
This resulted in approximately 65 million tweets, with approximately 2 to 2.5 million recorded per day.
We then extracted the accounts and all associated metadata (\eg account age) from tweets.
In total, we recorded information on 2.9 million unique accounts.
In this study, in addition to known metrics (age, tweets, retweets, favourites, replies and mentions, URL count, follower-friend ratio, \etc) we also analyse a set of six novel metrics not explored in past bot research.
These include: {\em likes per tweet}, {\em retweets per tweet}, {\em user replies and mentions}, {\em activity source count}, {\em type of activity sources}, and {\em size of content uploaded}.
The selection of features is driven by~\cite{dugue2015reliable} and, to our knowledge, this is the most comprehensive study to date.


\subsection{Data Pre-Processing}
\label{subsec:datapreprocessing}
Our data contains a range of accounts in terms of their popularity (\ie number of followers).
During preliminary analysis we found that the purpose and activity of an account differs based on its popularity.
Hence, we partition profiles into four popularity groups to enable a deeper understanding.
These are as follows:
\begin{itemize}
\item \gtenm {\bf -- celebrity status:}
This is the subset of Twitter users with the highest number of followers, \ie $>$9M followers.
These are the most popular users that hold celebrity status (\eg katyperry, BillGates) and are globally renowned (\eg CNN, NetGeo).
Popular and credible organisations use these accounts for various purposes, which makes them free of spam, thus having high credibility and trustworthiness.
\item \gonem {\bf -- very popular:}
This subset of Twitter users is amongst the most popular on the platform, \ie 900K to 1.1M followers.
These users are close to celebrity status (\eg jimcramer, BreeOlson) and global recognition (\eg nytfood, pcgamer).
\item \ghundredk {\bf -- mid-level recognition:} This subset represents popular accounts with mid-level recognition (\eg AimeeSayah, DavidJohnsonUK, CBSPhilly, DomusWeb), \ie 90k to 110k followers.
\item \gonek {\bf -- lower popularity:} This subset represents more ordinary users, \ie 0.9k to 1.1k followers.
These users (\eg hope\_bot, Taiwan\_Agent) form a large base and, though they show lower individual and accumulated activity, they do form the all-important tail of the distribution.
\end{itemize}

Our dataset is a representative sample of Twitter users, where each metric follows Gaussian distribution.
\gtenm and \gonem are similar in their characteristics (\cf \S\ref{sec:mannersmakethbot}) and constitute 0.65\% of the total 105k accounts we partitioned in the dataset.
\gonek represents the bulk of Twitter, constituting 94.40\% of the total partitioned accounts.
\ghundredk bridges the gap between the most popular and least popular groups, constituting 4.93\% of the total partitioned accounts.
A possible ${\bf G}_{\bf 10k}$ would be similar to \gonek, and a possible ${\bf G}_{\bf 50k}$ will be similar to \ghundredk.


\subsection{Bot Classification}
\label{subsec:botclassification}

To compare bots with humans, it is next necessary to identify which accounts are operated by bots.
We initially experimented with BotOrNot~\cite{Davis2016}, a state-of-the-art bot detection tool (to the best of our knowledge, this is the only available online tool).
However, inspection of the results indicated quite high levels of inaccuracy.
Hence, we chose to take a manual approach instead --- we made this design choice to have a smaller but more reliable set of classifications.
We employed human participants to perform a {\em human annotation task}\footnote{Human annotation task -- \scriptsize\texttt{http://bit.ly/2cH0YvA}} to identify bots and humans.
We note this could also serve as a reliable ground-truth dataset to train classification algorithms for automated bot detection in the future.


We recruited four undergraduate students for the purposes of annotation, who classified the accounts over the period of a month.
This was done using a tool that automatically presents Twitter profiles, and allows the recruits to annotate the profile with a classification (bot or human) and add any extra comments.
Each account was reviewed by all recruits, before being aggregated into a final judgement using a final collective review (via discussion among recruits if needed).

As well as providing the recruits with the Twitter profile, we also presented summary data to streamline the task. This included: account creation date, average tweet frequency, content posted on user Twitter page, account description, whether the user replies to tweets, likes or favourites received and the follower-friend ratio.
We also provide participants with a list of the `sources' used by the account over the month, \eg Twitter app, browser, \etc.
The human workers consider both the number of sources used, and the types of sources used.
This is because sources can reveal traces of automation, \eg use of the Twitter API.
Additionally, the human worker would also visit a user's Twitter page and verify the content and URLs posted.
Overall, we presented participants with randomised lists that fell into the four popularity groups containing 2500 accounts each.
Human annotators were instructed to filter out any account that matched the following criteria: account that does not exhibit activity (\ie no tweet, retweet, reply, and mention), or account that is suspended.
In total, the volunteers successfully annotated 3535 accounts.
Out of the successfully annotated accounts, 1525 were classified as bots and 2010 as humans.
At the time of writing all of the accounts in our dataset were active.
Table \ref{tab:datasummary} provides a summary of the data.


\begin{table}[t]
	\small
	\centering
	\caption{Summary of Twitter dataset post-annotation.}
	\begin{tabular}{|c|p{1.4cm}|p{1.3cm}|p{1.4cm}|p{1.3cm}|} \hline
	Group & \#Bot accounts & \#Human accounts & \#Bot statuses & \#Human statuses \\ \hline\hline
	\gtenm & 25 & 25 & 70713 & 77750 \\
	\gonem & 295 & 450 & 23447 & 25991 \\
	\ghundredk & 707 & 740 & 29777 & 21087 \\
	\gonek & 499 & 794 & 16112 & 5218 \\
	\hline
	{\bf Total} & 1525 & 2010 & 140049 & 130046 \\
	\hline
	\end{tabular}
\label{tab:datasummary}
\end{table}

For context, we can cross validate by comparing the agreement of final annotations by the human workers to the BotOrNot annotation.
The average inter-annotator agreement compares the pairs of labels by each human annotator to capture the percentage of accounts for which all four annotators unanimously agree.
The average agreement is measured as a percentage of agreement, where 0\% shows lack of agreement and 100\% shows perfect agreement.
Our human annotation task shows very high unanimous agreement between human annotators for each popularity group: \gtenm (96.00\%), \gonem (86.32\%), \ghundredk (80.66\%), and \gonek (93.35\%).
Whereas, BotOrNot shows lower than average agreement with the final labels assigned by the human annotators: \gtenm (46.00\%), \gonem (58.58\%), \ghundredk (42.98\%), and \gonek (44.00\%).
Since, BotOrNot yields a lower accuracy, we restricted ourselves to the dataset of accounts that were manually annotated.

\section{Which manners maketh the Bot?}
\label{sec:mannersmakethbot}


\begin{figure}[h]
\centering
\includegraphics[width=0.75\linewidth]{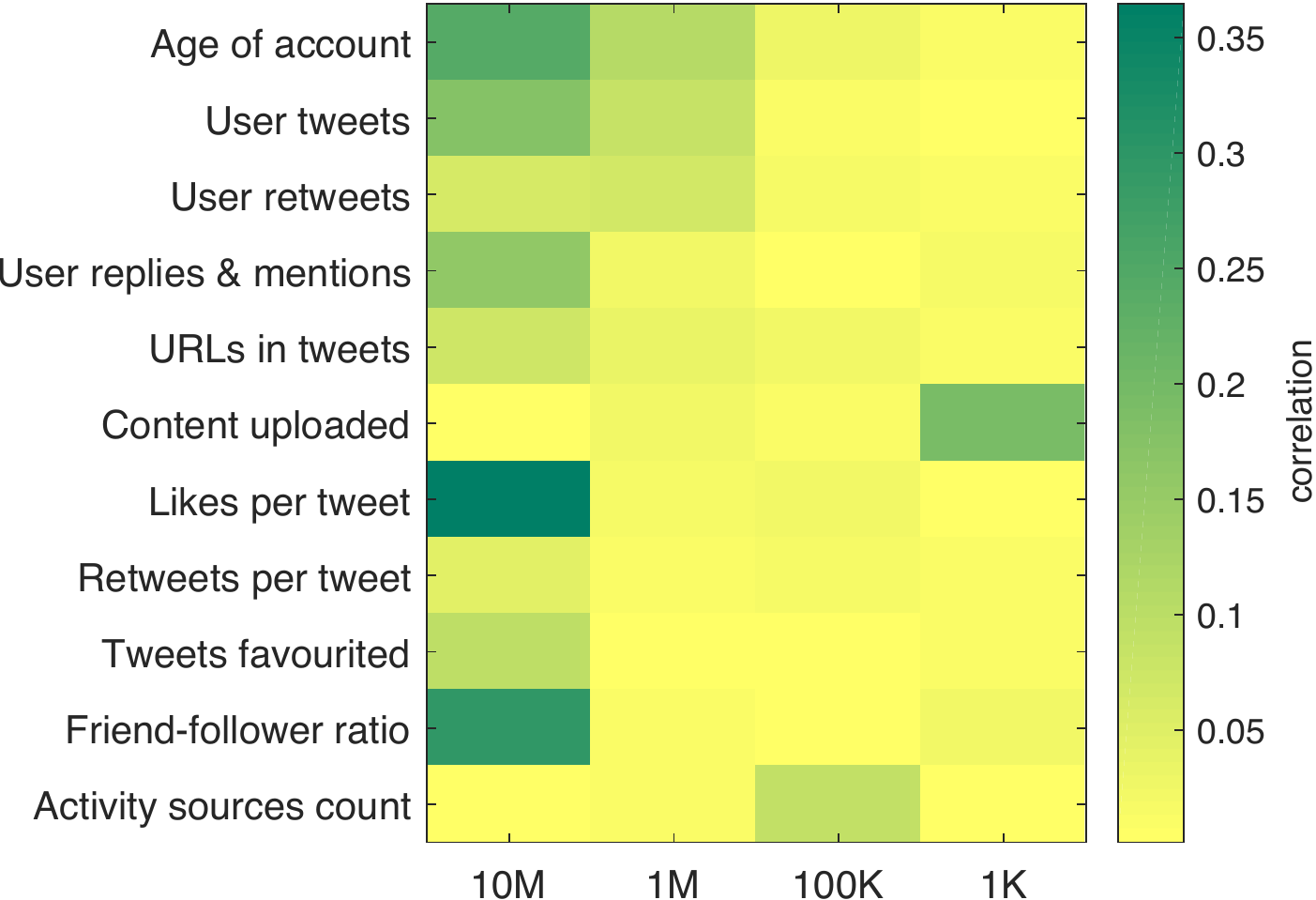}
\vspace{-0.2cm}
\caption{Spearman's rank correlation coefficient ($\rho$) between bots and humans per measured metric. The figure shows none (0.0) to weak correlation (0.35) across all metrics, indicating clear distinction between the two entities.}
\label{fig:feature_corr}
\vspace{-0.2cm}
\end{figure}

The purpose of this study is to discover the key account characteristics that are typical (or atypical) of bots and humans.
To explore this, we use our data (\S\ref{sec:methodology}) to empirically characterise bots (dashed lines in figures) and humans (solid lines in figures). 
To begin, we simply compute the correlation between each feature for bots and humans; Figure~\ref{fig:feature_corr} presents the results as a heatmap (where perfect correlation is 1.0). It can be immediately seen that most features exhibit very poor correlations (0.0 to 0.35), indicating significant discrepancies between bot and human behaviour --- we therefore spend the remainder of this paper exploring these differences in depth.





\subsection{Account Age}
\label{subsec:accountage}

\begin{figure}[h]
\centering
\includegraphics[width=0.75\linewidth]{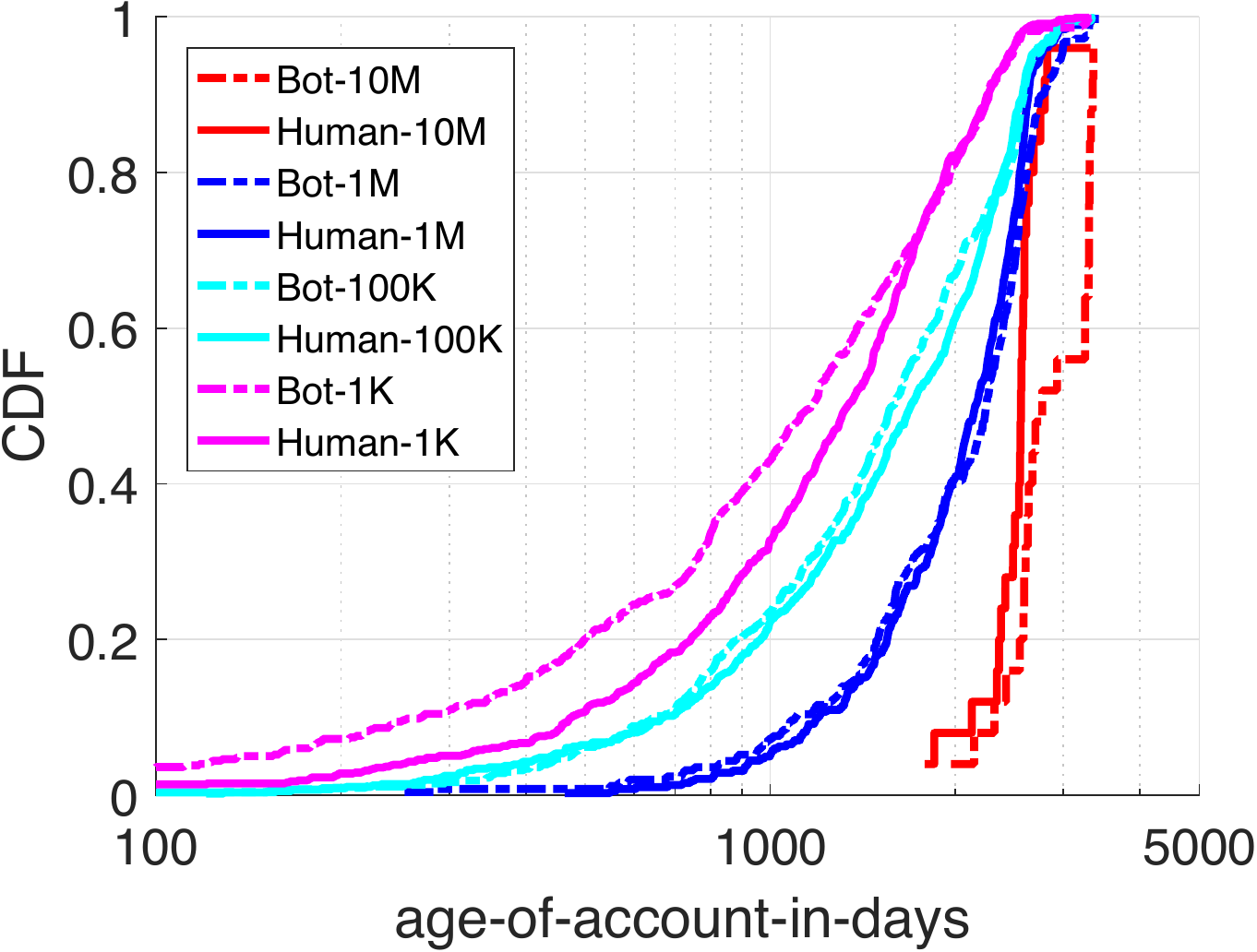}
\caption{Age of user account in days.}
\label{fig:age_of_account_in_days}
\end{figure}

First, we ask {\em how recently have bots started to participate on Twitter?}
To answer this we use account age, which captures the length of time an account has been in operation.
Figure \ref{fig:age_of_account_in_days} presents the age of each account, separated into their respective groups (bots \vs humans of varying popularities).
Unsurprisingly, it can be seen that there is a clear correlation between the age of an account and the number of followers.
More noteworthy is the fact that counterpart human accounts are actually slightly `younger' in age than the bot accounts, especially for \gtenm.
The oldest bot account is 3437 days old \vs 3429 days for the oldest human account -- bots are clearly not a new phenomenon.
Labels provided by human annotators suggest that these long-standing popular bot accounts have been operated by large reputable organisations as a means of reaching out to their fanbase, viewers and customers, \eg CNN, BBCWorld, NatGeo.
In particular, they are interested in spreading news, content, products and services, and are, thus, a very effective tool for these purposes.
This further confirms that the Twitter API incentivises the creation and existence of bots on its platform, as highlighted in \S \ref{sec:introduction}.
In fact, the age of some bot accounts suggest that they have long supported the evolution and uptake of Twitter.


\subsection{Content Generation}
\label{subsec:contentcreation}

\begin{figure*}[h]
\centering
	\subfigure[{Number of tweets issued by a user.}] {{\includegraphics[width=0.495\linewidth]{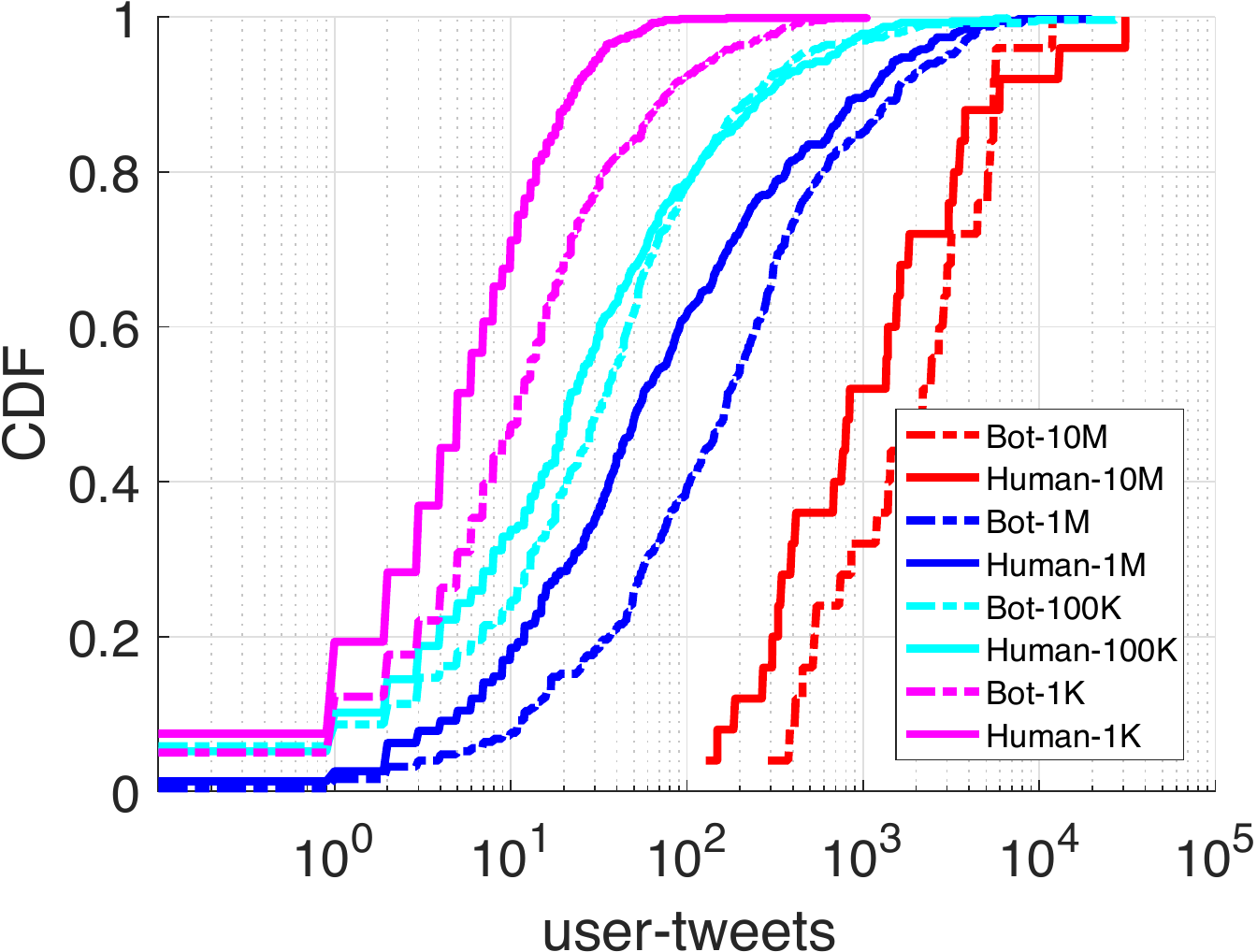}\label{fig:user_tweets}}}
	\subfigure[{Number of retweets issued by a user.}] {{\includegraphics[width=0.495\linewidth]{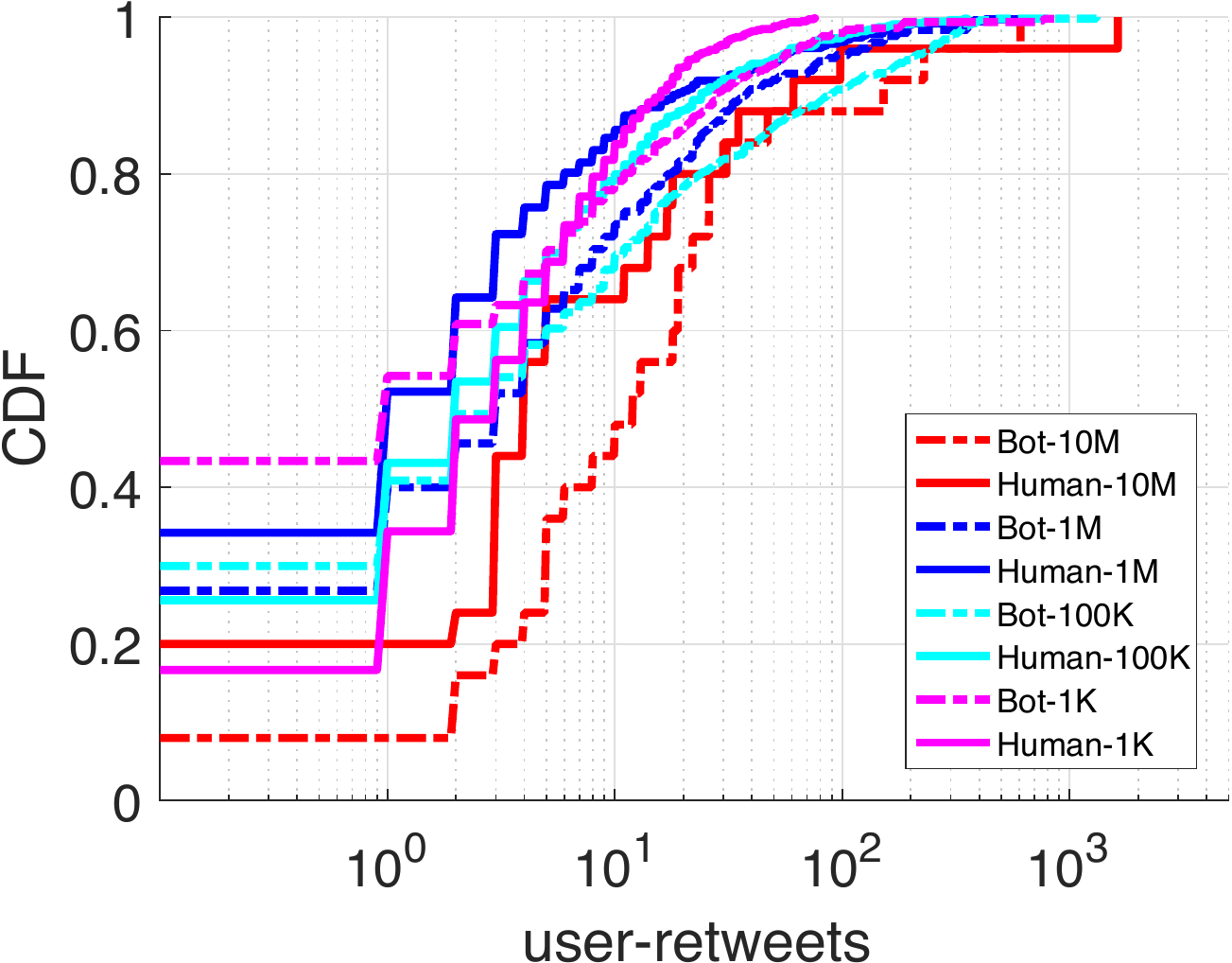}\label{fig:user_retweets}}}
\caption{Content Creation: Tweets issued, Retweets issued.}
\label{fig:manners_maketh_bot_1}
\end{figure*}

Next, we ask {\em if bots generate more content on Twitter than humans?}
Intuitively, one might imagine bots to be capable of generating more content, however, creativity is a clear bottleneck.
We initially consider two forms of content creation: a \emph{tweet}, which is an original status written by the account, and a \emph{retweet}, which is repetition of an existing status. When using the term \emph{status}, we are referring to the sum of both tweets and retweets.
First, we inspect the amount of content shared by computing the number of statuses (\ie tweets + retweets) generated by each account across the 30 days.
As anticipated, humans post statuses less frequently than bots (monthly average of 192 for humans \vs 303 for bots), in all popularity groups except \gtenm, where surprisingly humans post slightly more than bots.
The sheer bulk of statuses generated by \gtenm (on average 2852 for bots, 3161 for humans in a month) is likely to acquire popularity and new followers.
Overall, bots constitute 51.85\% of all statuses in our dataset, even though they are only 43.14\% of the accounts.
An obvious follow-up is {\em what do accounts tweet?} This is particularly pertinent as bots are often reputed to lack original content.
To explore this, we inspect the number of {\em tweets} \vs ~{\em retweets} performed by each account.
Recall, that a tweet is an original status and not a retweet.
Figures \ref{fig:user_tweets} and \ref{fig:user_retweets} present the empirical distributions of tweets and retweets, respectively, over the 30 days.
We see that the retweet distribution is rather different to tweets.
Bots in \gonem, \ghundredk and \gonek are far more aggressive in their retweeting; on average, bots generate 2.20$\times$ more retweets than humans.
The only exception to this trend is \gtenm where humans retweet 1.54$\times$ more often than bots.
This is likely driven by the large number of tweets generated by celebrity users --- clearly, when tweeting over a hundred times a day, it becomes necessary to rely more heavily on retweeting.
That said, typically, humans do generate {\em new} tweets more often, while bots rely more heavily on retweeting existing content.
Generally, humans post 18 tweets for every retweet, whereas bots post 13 tweets for every retweet in all popularity groups except \gtenm (where both entities show similar trends).

%

\begin{figure}[h]
\centering
\includegraphics[width=0.75\linewidth]{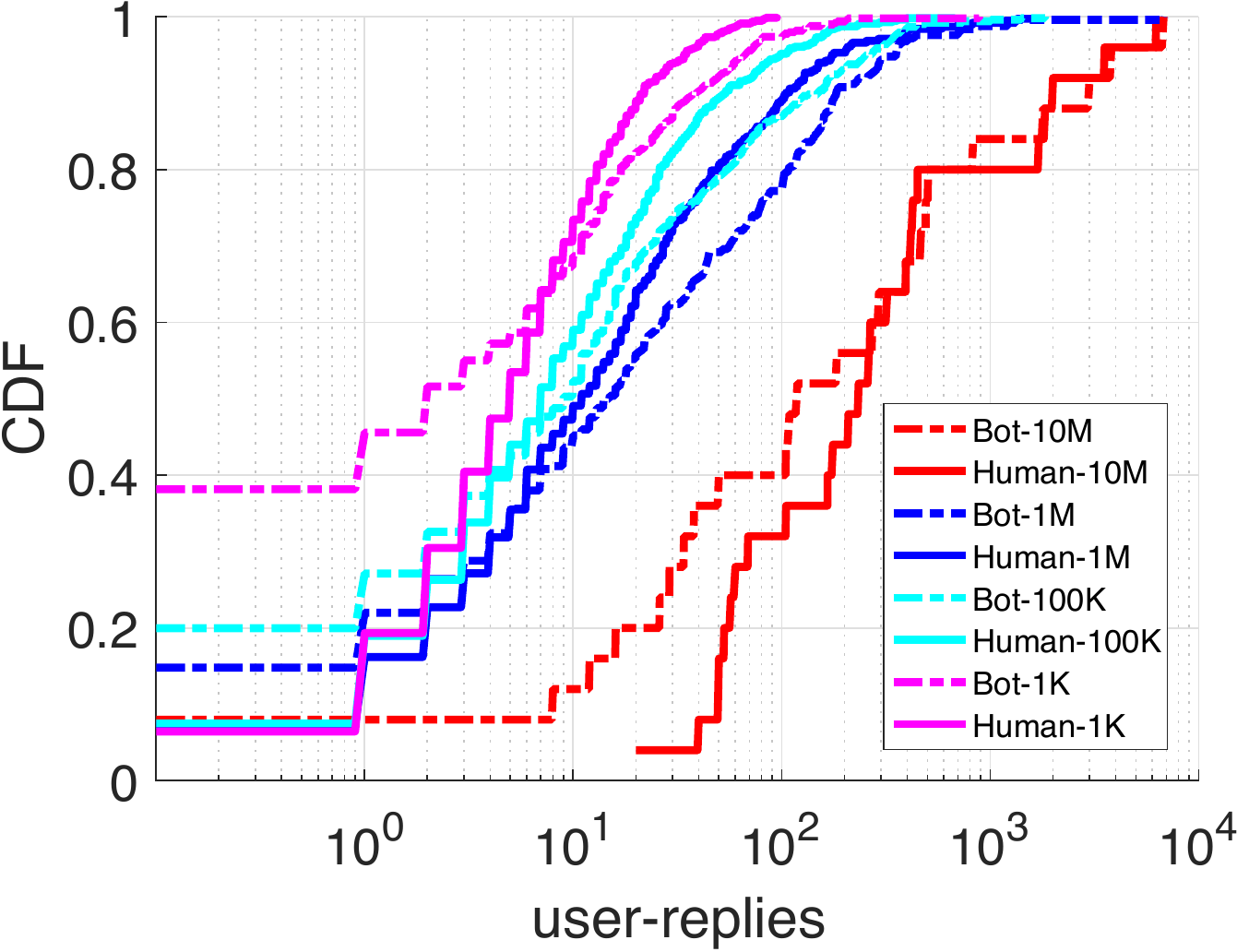}
\caption{Content Creation: Number of replies and mentions posted by a user.}
\label{fig:user_replies_and_mentions}
\end{figure}

Whereas tweets and retweets do not require one-to-one interaction, a further type of messaging on Twitter, via {\em replies}, does require one-to-one interaction.
These are tweets that are created in response to a prior tweet (using the @ notation).
Figure \ref{fig:user_replies_and_mentions} presents the distribution of the number of replies issued by each account.
We anticipate that bots post more replies and mentions given their automated capacity to do so.
For \gtenm, bots post only marginally more, on average, than celebrities.
This appears to be due to two reasons: bots in \gtenm deploy {\em chatbots} for addressing simple user queries, and need-based human intervention to engage with followers.
From the perspective of celebrities, replies serve the need (or desirability) to interact with their fanbase.
For the most popular automated accounts it is the need to promote direct interaction for a larger user base.
Bots in the remaining popularity groups respond twice as frequently as their human counterparts.
Again, this seems to be driven by the ease by which bots can automatically generate replies: only the most dedicated human users can compete.

\begin{figure*}[t]
\centering
	\subfigure[{Content uploading activity -- URLs \vs Content uploaded.}] {{\includegraphics[width=0.495\linewidth]{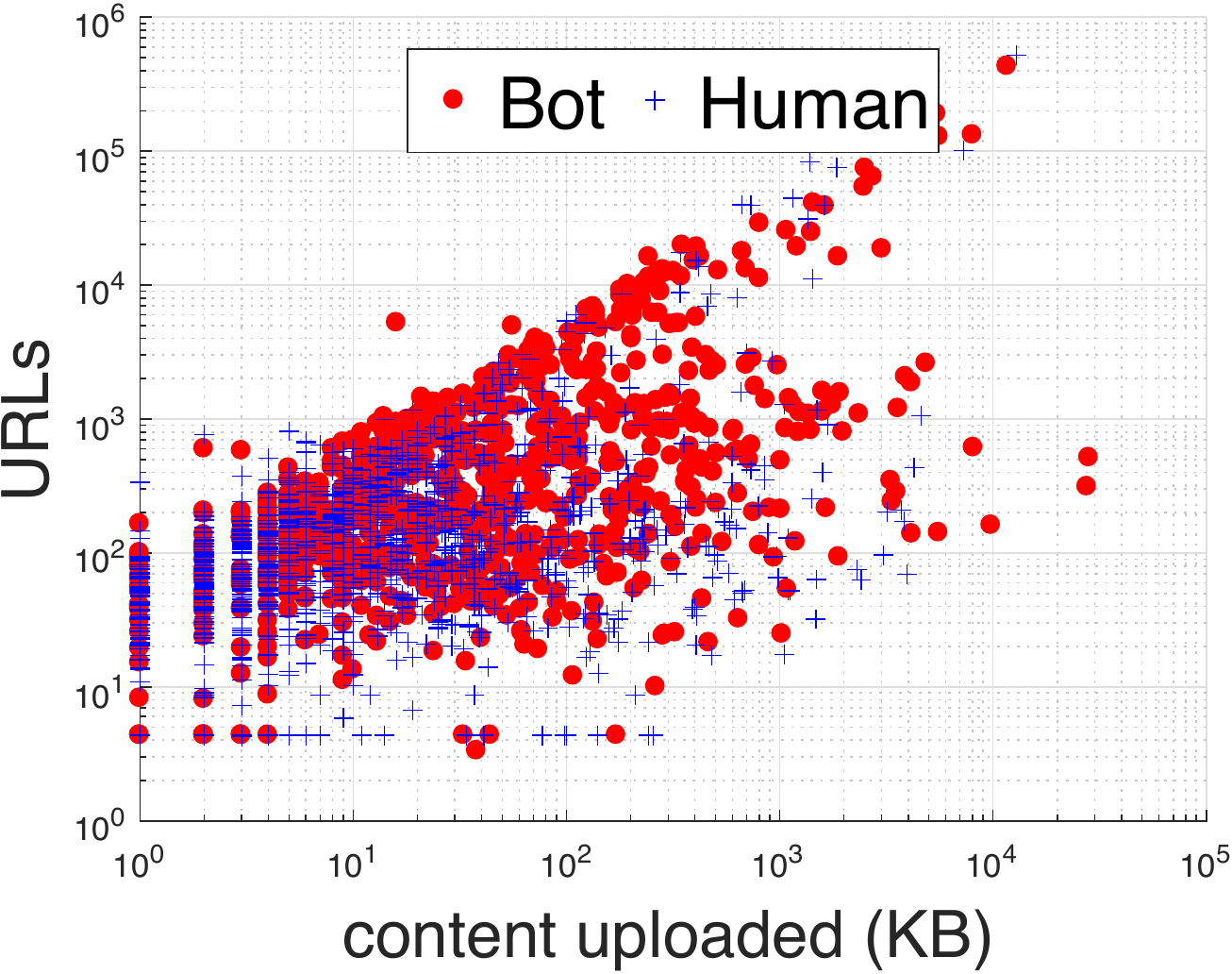}\label{fig:url_content_uploading}}}
	\subfigure[{Size of CDN content in KByte uploaded by a user.}] {{\includegraphics[width=0.495\linewidth]{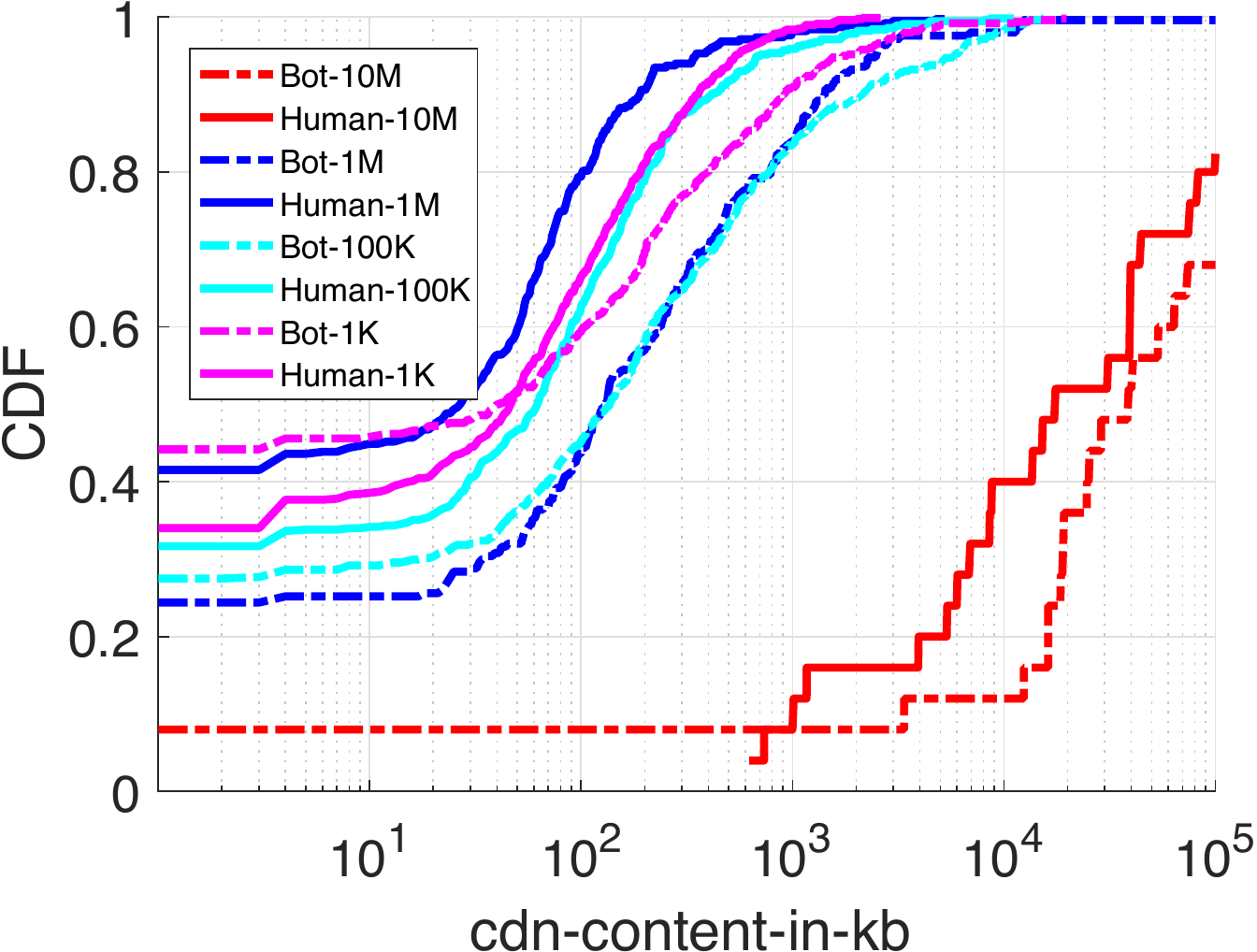}\label{fig:content_size}}}
\caption{Content Creation: URLs in tweets, Content uploaded on Twitter.}
\label{fig:manners_maketh_bot_2}
\end{figure*}

Finally, we briefly inspect the actual content of the tweets being generated by the accounts.
We do this using two metrics: number of URLs posted by accounts, and the size of media (\eg pictures) uploaded.
Figure \ref{fig:url_content_uploading} presents the scatter plot of the number of URLs ($y$-axis) and content uploaded in KB ($x$-axis).
Bots place far more external URLs in their tweets than humans (see Table \ref{tab:featureinclination}): 162\% in \gtenm, 206\% more in \gonem, 333\% more in \ghundredk, and 485\% more in \gonek.
It has been previously noted that a higher number of external URLs suggests automation \cite{Chu2010}.
Bots are a clear driving force for generating traffic to third party sites, and upload far more content on Twitter than humans.
Figure \ref{fig:content_size} presents the distribution of the amount of content uploaded by accounts (\eg images). 
This is extracted by finding matches for \texttt{\small *.twimg.*} keyword in the \texttt{\small [entities] [media] [media\_url]} tweet attribute.
Note that \texttt{\small *.twimg.*} is a generic domain name for Twitter CDN\footnote{What is \texttt{twimg}? -- \scriptsize\texttt{http://bit.ly/2mCnSL0}}.
Account popularity has a major impact on this metric.
Bots in \gtenm have a 102$\times$ lead over bots in other popularity groups.
That said, humans in \gtenm have a 366$\times$ lead over humans in other popularity groups.
We conjecture that celebrities have a lot more to share from their lives, but also use uploading as a means to stay connected with their fanbase.
Overall, bots upload substantially more bytes than humans do (see Table \ref{tab:featureinclination}): 141\% in \gtenm, 975\% more in \gonem, 376\% more in \ghundredk, and 328\% more in \gonek.
It is worth noting that both content upload and URL inclusion trends are quite similar, suggesting that both are used with the same intention, \ie spreading content.
Since bots in \gtenm mostly belong to news media -- sharing news headlines is clearly a means of operating their business.

%

\subsection{Content Popularity}
\label{subsec:contentpopularity}

The previous section has explored the amount of content generated by accounts, however, this does not preclude such content from being of a low quality.
To investigate this, we compute standard popularity metrics for each user group.

\begin{figure*}[h]
\centering
	\subfigure[{Likes per tweet received by a user.}] {{\includegraphics[width=0.495\linewidth]{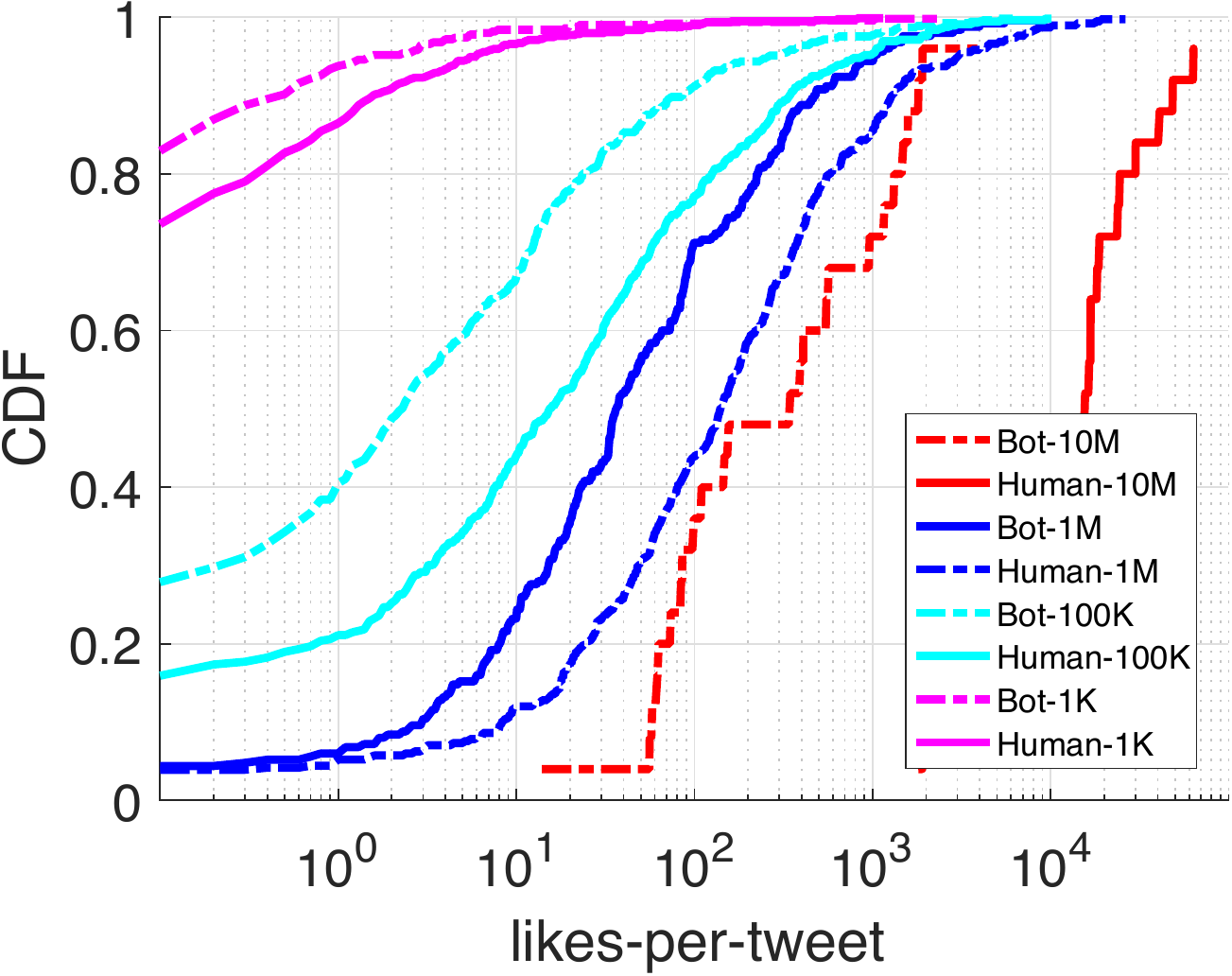}\label{fig:likes_per_tweet}}}
	\subfigure[{Retweets per tweet received by a user.}] {{\includegraphics[width=0.495\linewidth]{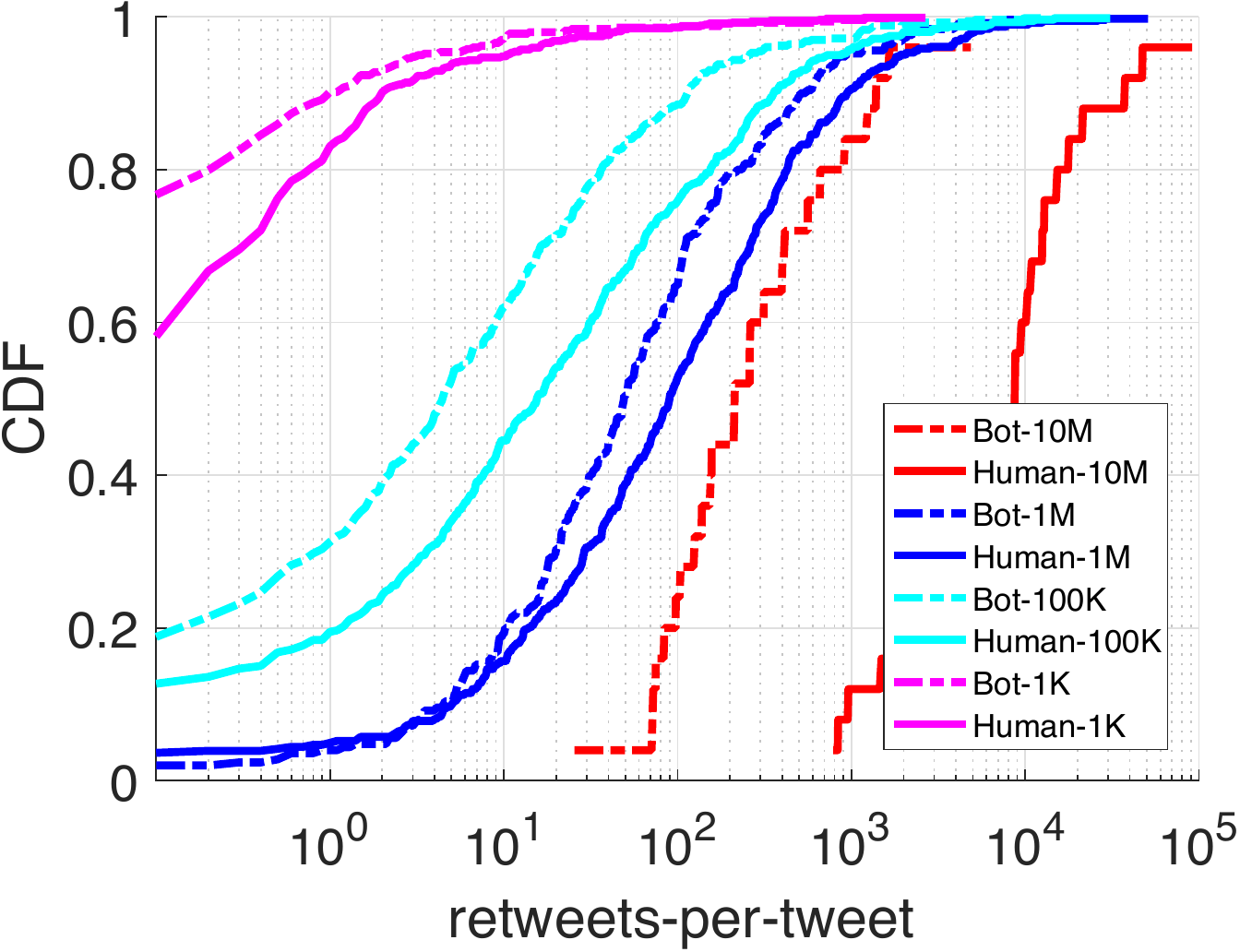}\label{fig:retweets_per_tweet}}}
\caption{Content Popularity: Likes per tweet, Retweets per tweet.}
\label{fig:manners_maketh_bot_3a}
\end{figure*}

First, we inspect the {\em number of favourites} or {\em likes} received for tweets generated by the accounts.
This is a reasonable proxy for tweet quality.
Figure \ref{fig:likes_per_tweet} presents the empirical distribution of the number of favourites or likes received for all the tweets generated by the profiles in each group.
A significant discrepancy can be observed.
Humans receive {\em far} more favourites per tweet than bots across all popularity groups except \gonek.
Close inspection by human annotators revealed that bots in \gonek are typically part of larger {\em social botnets} that try to promote each other systematically for purposes as outlined in \S \ref{sec:introduction}, whereas human accounts are limited to their social peers and do not usually indulge in the `influence' race.
For \gtenm, \gonem and \ghundredk popularity groups, humans receive an average of 27$\times$, 3$\times$ and 2$\times$ more favourites per tweet than bots, respectively.
\gonek bots are an exception that receive 1.5$\times$ more favourites per tweet than humans.
This is a clear sign that the term {\em popularity} may not be ideally defined by the number of followers.
According to the human annotators, this is due to a number of reasons including: humans posting personal news, engaging with their followers, posting less often than bots, staying relevant, posting for their fans (personal things from their life), and not aiming to redirect traffic to external websites.
Bots on the other hand, post much more often and do not engage with followers, likely reducing the chances of their tweets being liked.

\begin{figure*}[h]
\centering
	\subfigure[{Interaction graph for \gtenm. Black dots are accounts, edges are interactions. \textcolor{red}{Red} edges are bots; \textcolor{blue}{Blue} edges are humans.}] {{\includegraphics[width=0.495\linewidth]{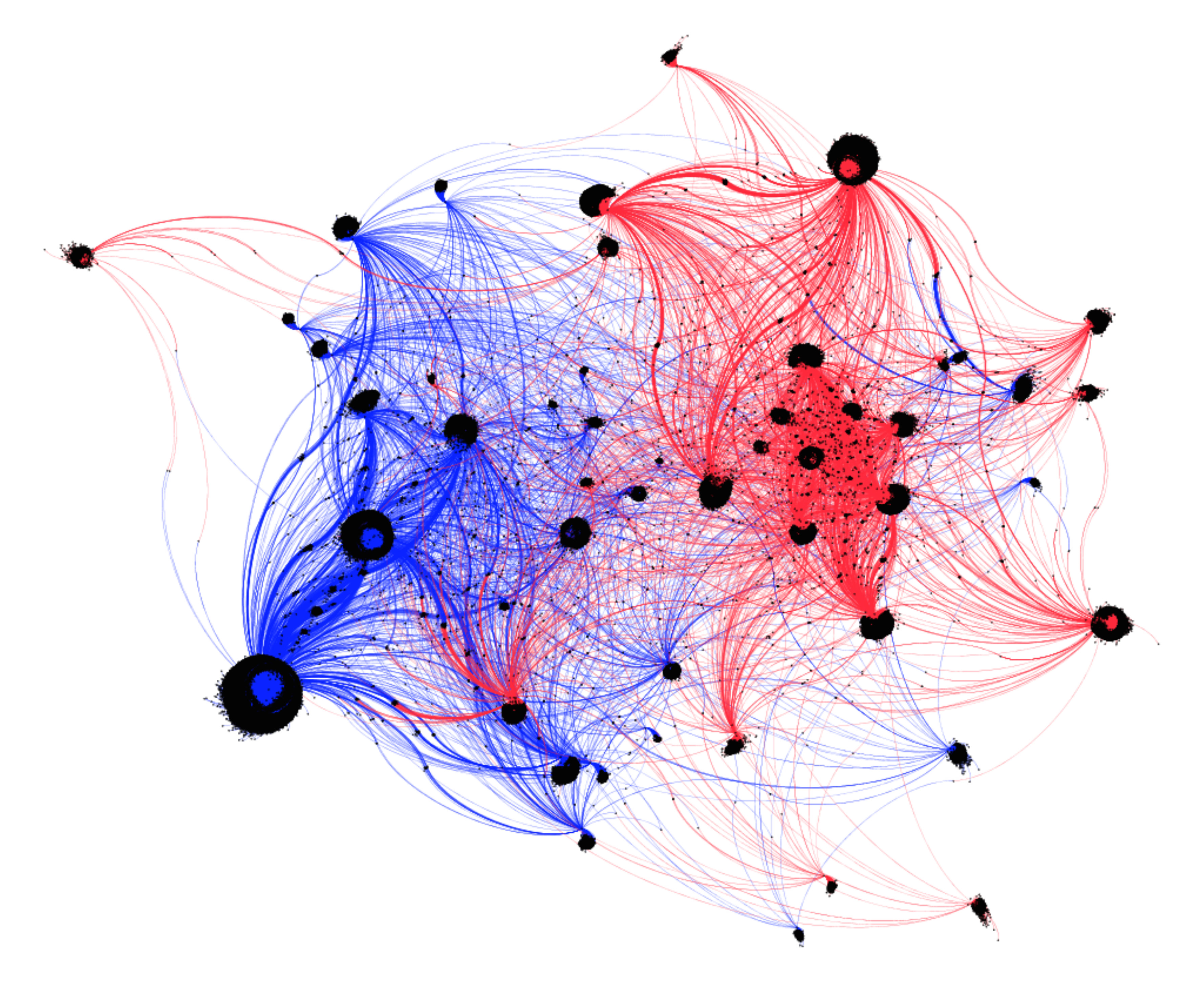}\label{fig:10M_all}}}
	\subfigure[{\gtenm Humans only (Bots removed).}] {{\includegraphics[width=0.495\linewidth]{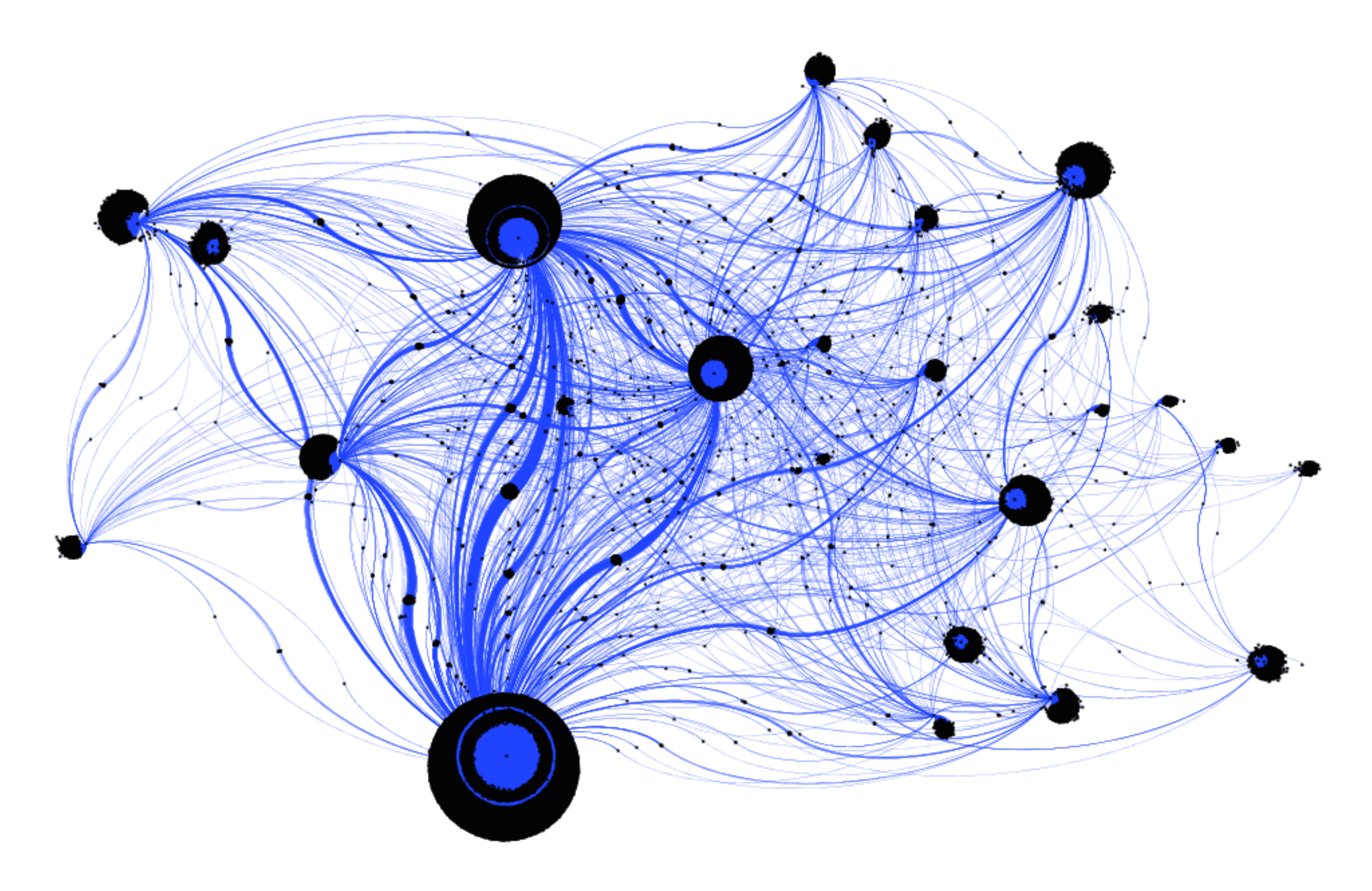}\label{fig:10M_humans}}}
\caption{Content Popularity: Interaction graphs for Bots \vs Humans.}
\label{fig:manners_maketh_bot_3b}
\end{figure*}

A further sign of content quality is another user retweeting content.
This is potentially an even stronger signal of endorsement, as a retweet will explicitly be listed on a user's wall.
Humans consistently receive more retweets for all popularity groups \gtenm: 24-to-1, \gonem and \ghundredk: 2-to-1, except \gonek: 1-to-1.
This difference, shown in Figure \ref{fig:retweets_per_tweet}, is indicative of the fanbase loyalty, which is vastly higher for individual celebrities than reputable organisations.
In other words, the {\em quality} of human content appears to be much higher.
We can then inspect \emph{who} performs the retweets, \ie do bots tend to retweets other bots or humans?
We find that bots retweeting other bots is over 3$\times$ greater than bots retweeting humans.
Similarly, humans retweeting other humans is over 2$\times$ greater than humans retweeting bots.
Overall, bots are retweeted 1.5$\times$ more than humans.
This indicates a form of homophily and assortativity.
To visualise this, Figure \ref{fig:10M_all} presents a graph of interactions for the 10M accounts; nodes are the accounts, whilst edges represent a direct interaction \ie retweeted statuses, quoted statuses, replies, and mentions.
The above observations of homophily can immediately be seen with bots and humans forming distinct groups.
To further confirm this, Figure \ref{fig:10M_humans} shows the status of interactions if bots are removed from the bot-human interactions (Figure \ref{fig:10M_all}).
Although interactions are still substantial, they are notably less when bots are excluded.
This indicates that, where popularity does exist, it tends to be driven by assortative account types, \eg popular bot accounts are given that status via other bots.


\subsection{Content Consumption}
\label{subsec:contentconsumption}

\begin{figure*}[h]
\centering
	\subfigure[{Tweets favourited (liked) by a user.}] {{\includegraphics[width=0.495\linewidth]{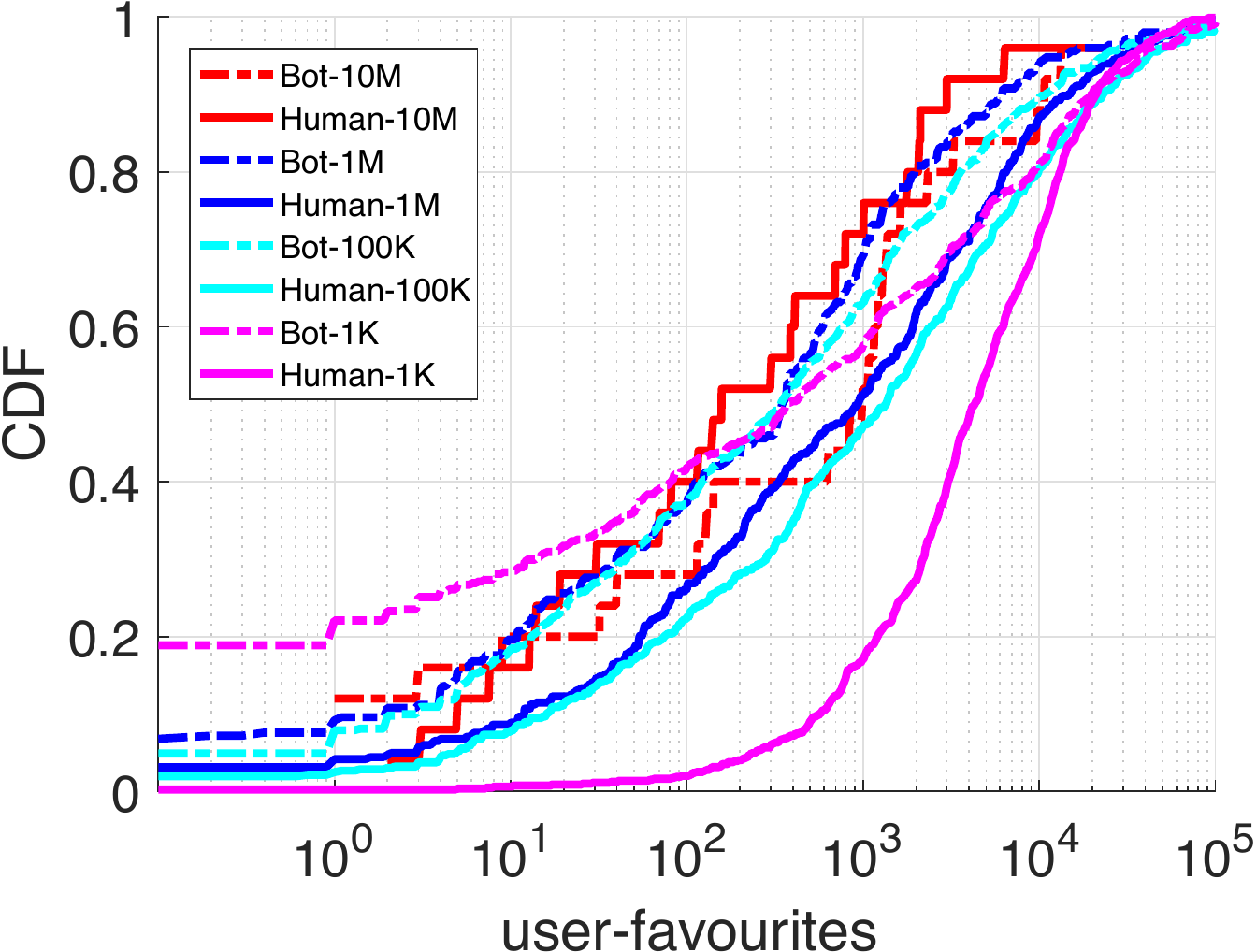}\label{fig:favourited_by_user}}}
	\subfigure[{Number of favourites performed \vs age of the account.}] {{\includegraphics[width=0.495\linewidth]{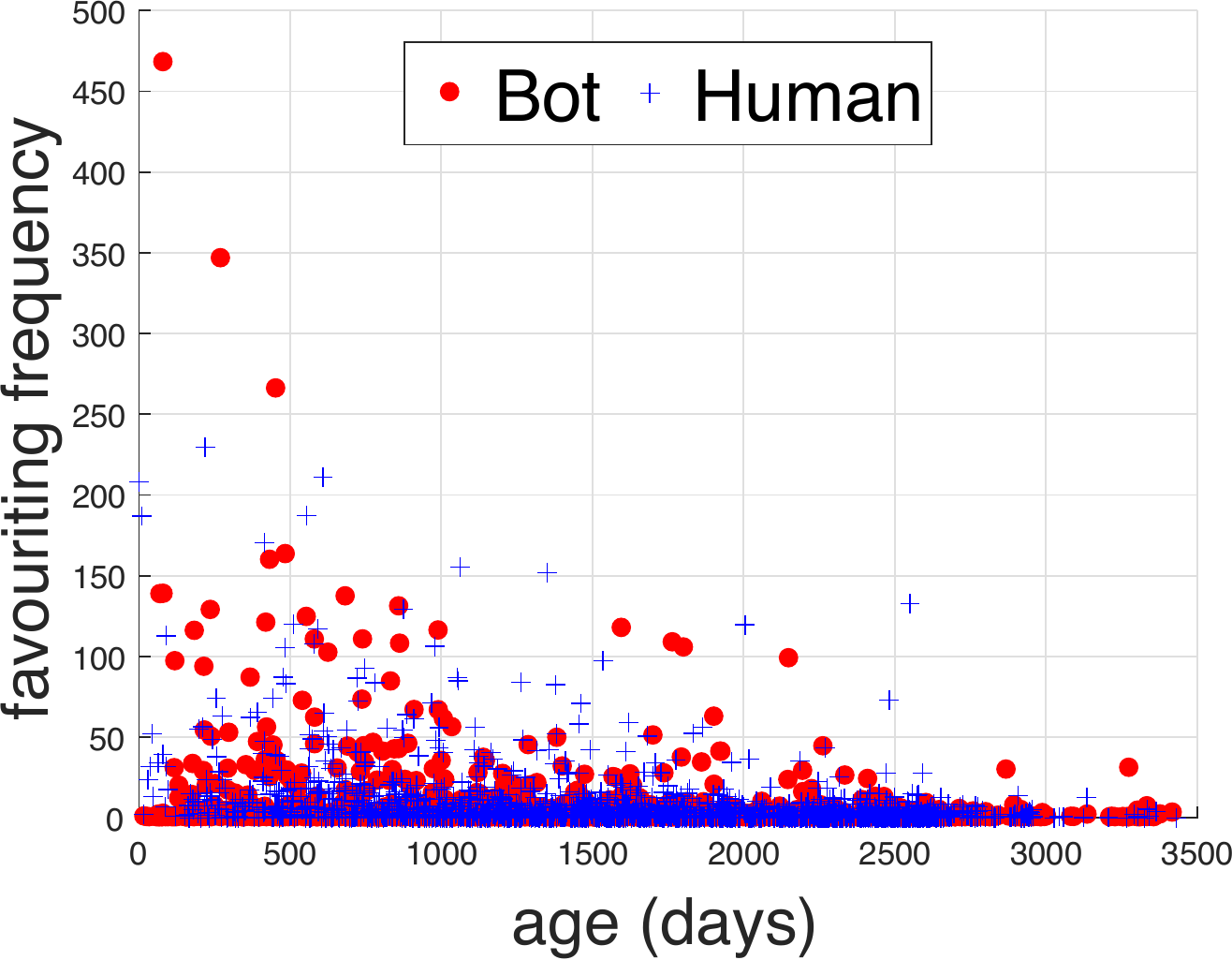}\label{fig:aggressive_bots}}}
\caption{Content Consumption: Likes performed, Favouriting behaviour.}
\label{fig:manners_maketh_bot_4}
\end{figure*}

Whereas the previous metrics have been based on content produced {\em by} the accounts under-study, our dataset also includes the consumption preferences of the accounts themselves.
Hence, we ask {\em how often do bots `favourite' content from other users and how do they compare to humans?}
Intuitively, bots would be able to perform far more likes than humans (who are physically constrained).
Figure \ref{fig:favourited_by_user} shows empirical distribution of the number of likes performed by each account.
It can be seen that, actually, for most popularity groups (\gonem, \ghundredk, \gonek), humans favourite tweets more often than bots (on average 8251 for humans \vs 5445 for bots across the entire account lifetimes).
Linking into the previous discussion, it therefore seems that bots rely more heavily on retweeting to interact with content.
In some cases, the difference is significant; for example, \gonem and \ghundredk see, on average, 2$\times$ more likes by humans compared to bots.
\gtenm, however, breaks this trend with an average of 1816 by humans compared to 2921 by bots.


We conjecture that there are several reasons for this trend.
First, humans have a tendency to appreciate what they like and therefore marking a like is a manifestation of that tendency.
In contrast, many bots are simply trying to promote their own tweets and therefore their interactions are based less on likes.
We also posit that humans have an incentive to like other tweets, potentially as a social practice (with friends) or in the hope of receiving likes in return \cite{Scissors:2016:WLA:2818048.2820066}.
Bots, on the other hand, are less discerning, and feel fewer social obligations.
To explore these strategies further, Figure \ref{fig:aggressive_bots} plots the number of favourites performed by an account \vs the age of the account. 
It can be seen that more recent (\ie more modern) bots are significantly more aggressive in liking other tweets.
Older bots, instead, use this feature less frequently; manual inspection suggests this is driven by the trustworthy nature of older bots, which are largely run by major organisations.
In contrast, younger bots adhere less to good social practice and clearly attempt to use favouriting to garner a lot of attention.

\subsection{Account Reciprocity}
\label{subsec:accountreciprocity}

\begin{figure}[h]
\centering
\includegraphics[width=0.75\linewidth]{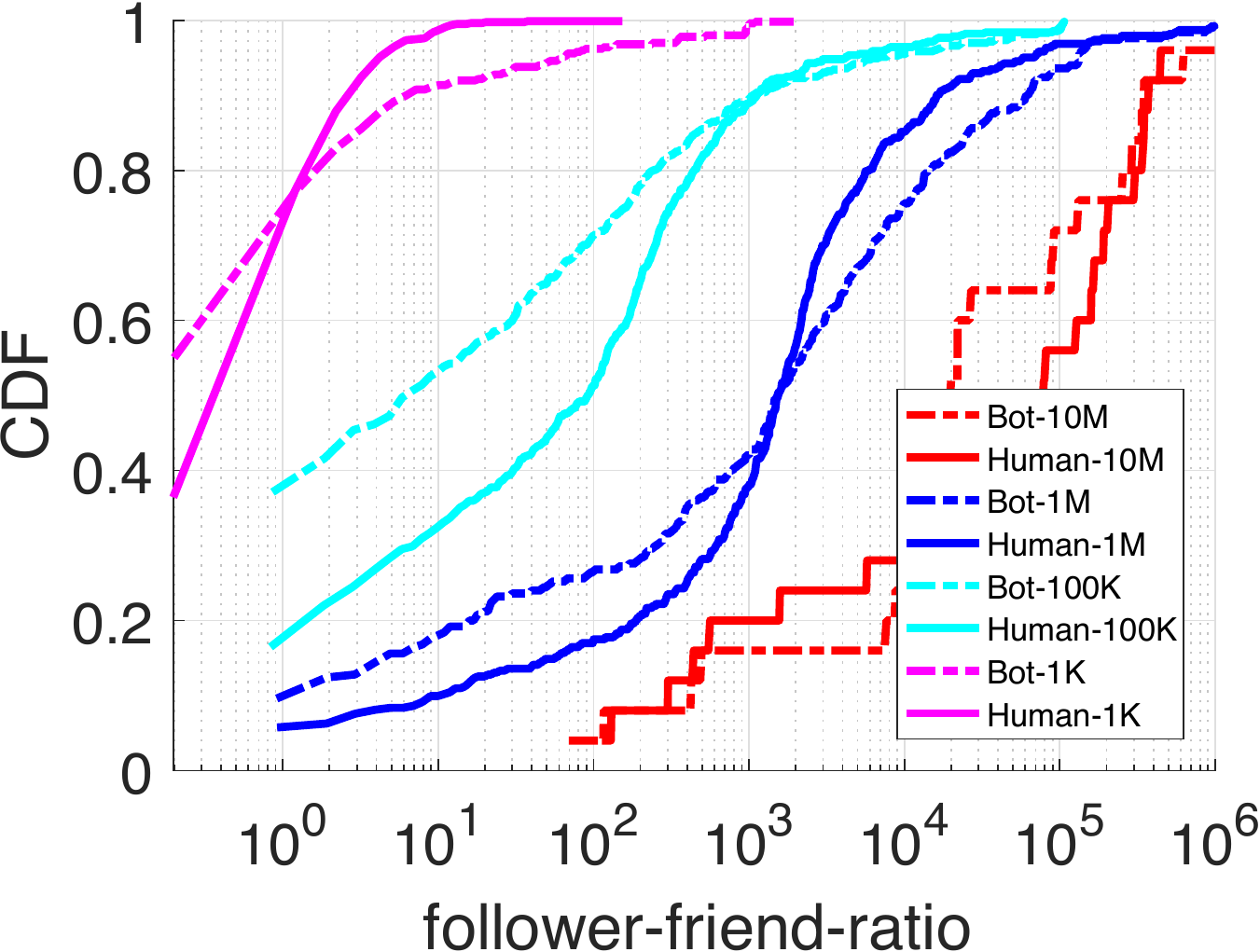}
\caption{Friend-follower ratio of a user.}
\label{fig:follower_friend_ratio}
\end{figure}

As well as content popularity, we can also measure reciprocity (\ie friendship).
In Twitter, this is classified as a reciprocal follower-relationship (\ie when two accounts follow each other).
This is in contrast to a non-reciprocal relationship, whereby an account has many followers who are not followed in return (this is often the case for celebrities).
We ask {\em do bots show reciprocity similar to humans?}
We measure this via the {\em Follower-Friend Ratio}.
Figure \ref{fig:follower_friend_ratio} shows empirical distribution of the {\em Follower-Friend Ratio} for each group of accounts.
The results show that humans display higher levels of friendship (\gtenm: 4.4$\times$, \gonem and \ghundredk: 1.33$\times$, \gonek: 15$\times$) and thus a lower {\em Follower-Friend Ratio} than bots.

Previous research \cite{Chu2010} argues that humans typically have a ratio close to 1, however, our analysis contradicts this assumption.
For celebrities, very popular and mid-level recognition accounts this ratio is in the order of thousands-to-1, irrespective of whether an account is a bot or a human (\gtenm: 629011-to-1 for bots \vs 144612-to-1 for humans, \gonem: 33062-to-1 for bots \vs 24623-to-1 for humans, \ghundredk: 2906-to-1 for bots \vs 2328-to-1 for humans).
In fact, even the ratios for low popularity accounts are not 1, but consistently greater (\gonek: 30-to-1 for bots \vs 2-to-1 for humans).
This is caused by a human propensity to follow celebrity accounts (who do not follow in return), as well as the propensity of bots to indiscriminately follow large numbers of other accounts (largely in the hope of being followed in return).

\subsection{Tweet Generation Sources}
\label{subsec:tweetgenerationsources}

\begin{figure*}[h]
\centering
	\subfigure[{Activity sources used by a user (\textcolor{red}{Red} dot is the mean of the distribution).}] {{\includegraphics[width=0.495\linewidth]{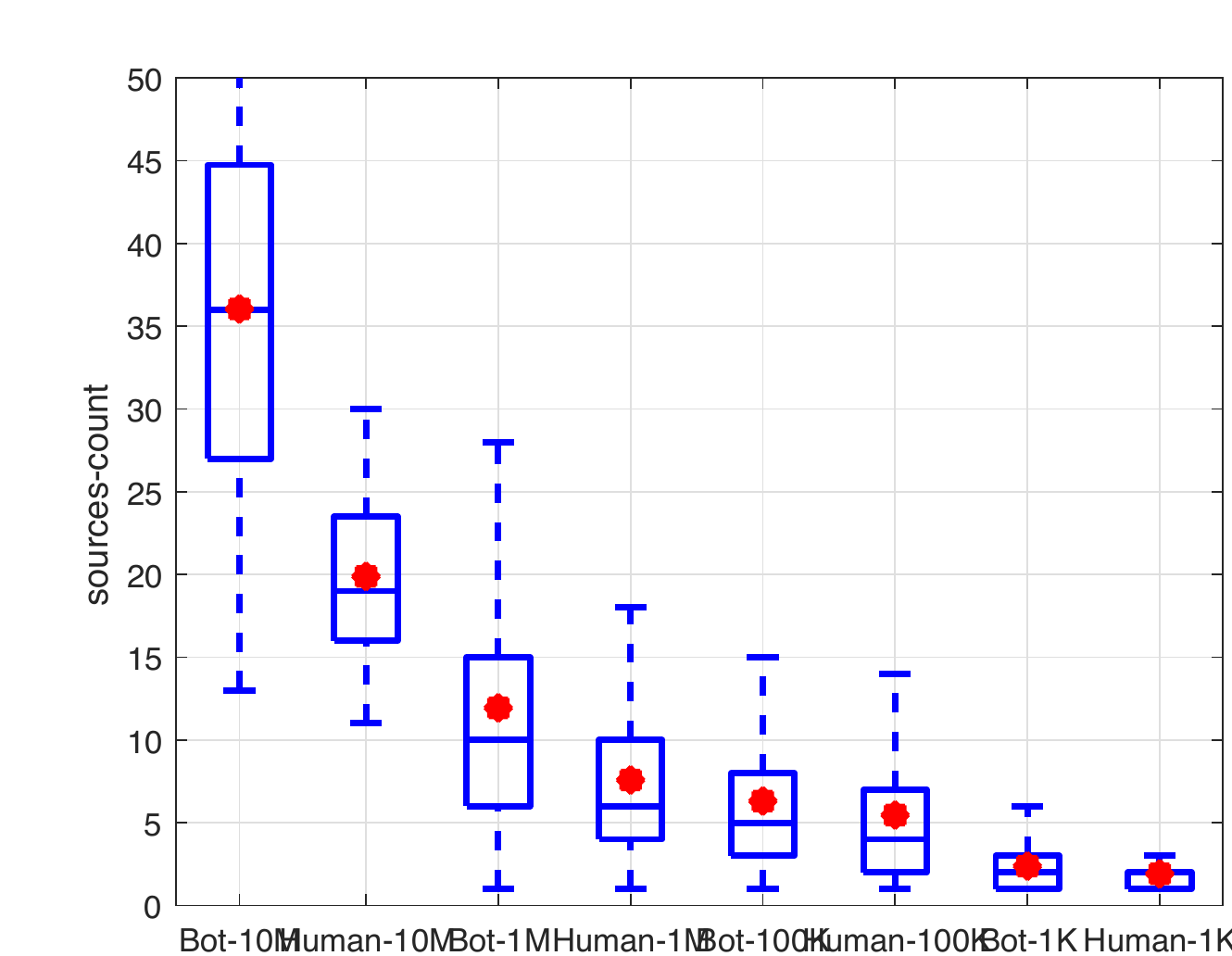}\label{fig:sources_count}}}
	\subfigure[{Bar chart of number of accounts that use each type of Twitter source.}] {{\includegraphics[width=0.495\linewidth]{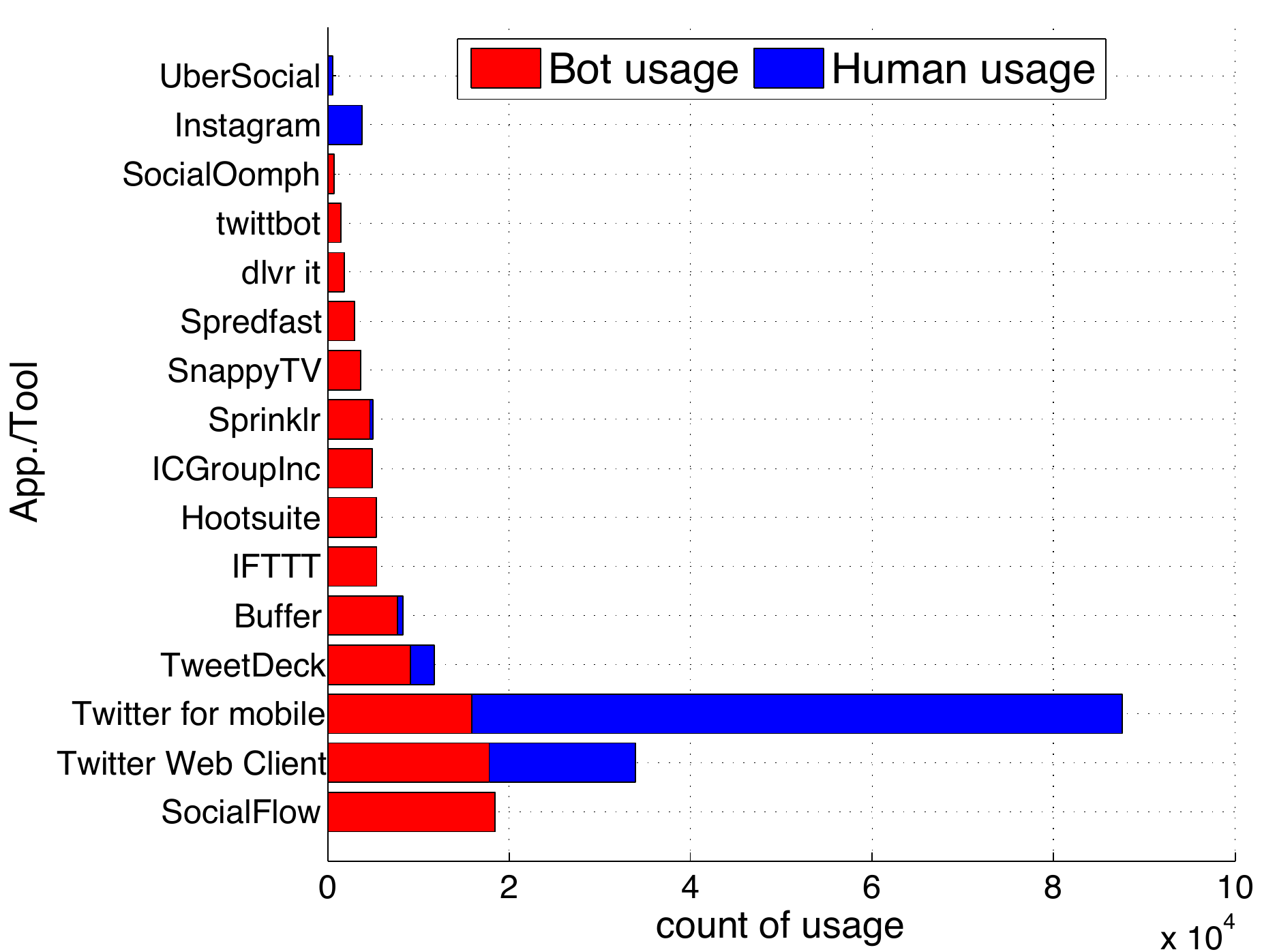}\label{fig:sources_types}}}
\caption{Tweet Generation Sources: Count of Activity Sources, Type of Activity Sources.}
\label{fig:manners_maketh_bot_5}
\end{figure*}

Finally, we inspect the tools used by bots and humans to interact with the Twitter service. This is possible because each tweet is tagged with the {\em source} that generated it; this might be the website, app or tools that employ the Twitter API.
Figure \ref{fig:sources_count} presents the number of sources used by human and bot accounts of varying popularities.
Surprisingly, it can be seen that bots actually inject tweets using more sources than humans (\cf Table \ref{tab:featureinclination}).
This is unexpected as one might expect bots to use a single source (\ie an API or their own automated tool).
However, paid third party services exist that can be utilised for the automation purpose.

%

To explore this further, Figure \ref{fig:sources_types} presents the number of accounts that use each source observed.
It can be seen that bots use a multitude of third-party tools.
Bot news services (especially from \gtenm) are found to be the heaviest users of social media automation management and scheduling services {\em (SocialFlow, Hootsuite, Sprinklr, Spredfast)}, as well as a Cloud-based service that helps live video editing and sharing {\em (SnappyTV)}.
Some simpler bots (from \ghundredk and \gonek groups) use basic automation services {\em (Dlvr.it, Twittbot)}, as well as services that post tweets by detecting activity on other platforms {\em (IFTTT)}.
We also detected use of social media marketing and branding services {\em (ICGroupInc, IFTTT, Spredfast)} by users (especially organisations operating user accounts via automated agents) in all popularity groups.
A social media dashboard management tool seems to be popular across most groups except \gonek {\em (TweetDeck)}.
Interestingly, it can also be seen that bot accounts regularly tweet using the Web/mobile clients --- it seems likely that these accounts have a {\em mix} of automated and human operation.
In contrast, 91.77\% of humans rely exclusively on the Web/mobile clients.
That said, a small number (3.67\%) also use a popular social media dashboard management tool {\em (TweetDeck)}, and automation and scheduling services {\em (Buffer, Sprinklr)}. 
This is particularly the case for celebrities, who likely use the tools to maintain high activity and follower interaction --- this helps explain the capacity of celebrities to so regularly reply to fans (\S\ref{subsec:contentcreation}).
These things should be noted when using the tweet sources to identify bots.

\section{Discussion, Conclusions \& Future Work}
\label{sec:discussion}

Bots exercise a profound impact on Twitter, as this paper thoroughly investigates.
Table \ref{tab:featureinclination} summarises key differences between bots and humans.
For each popularity group the table shows which feature most suitably distinguishes which entity, \ie bot ($\mathcal{B}$) or human ($\mathcal{H}$).
The magnitude of differences is represented by * (considerable) and ** (large).
Table \ref{tab:featureinclination} also includes statistical significance ({\em signif.}) of the mean \emph{difference} between humans and bots across all popularity groups, as measured by the {\em t-test}: at 99\% confidence level (extremely different), and 95\% (very different).

\begin{table}[t]
	\small
	\centering
	\caption{Feature inclination: $\mathcal{B}$ is more indicative of bots, whereas $\mathcal{H}$ is more indicative of human behaviour, and $\bigcirc$ is neutral (\ie both exhibit similar behaviour). * represents magnitude of inclination: * is considerable difference, ** is large difference. {\em signif.} shows statistical significance of each feature as measured by {\em t-test}.}
	\begin{tabular}{|p{4cm}|c|c|c|c|c|c|}\hline
	{\bf Feature \& value} & {\bf Fig.} & {\bf 10M+} & {\bf 1M} & {\bf 100K} & {\bf 1K} & {\bf \em signif.}\\ \hline
	{\em Older} age of account in days & \ref{fig:age_of_account_in_days} & $\mathcal{B}$ & $\bigcirc$ & $\bigcirc$ & $\bigcirc$ &  \\ \hline
	{\em More} user tweets & \ref{fig:user_tweets} & $\bigcirc$ & $\mathcal{B}$* & $\mathcal{B}$* & $\mathcal{B}$* &  \\ \hline
	{\em Higher} user retweets & \ref{fig:user_retweets} & $\mathcal{H}$* & $\mathcal{B}*$ & $\mathcal{B}*$ & $\mathcal{B}*$ & 99\% \\ \hline
	{\em More} user replies and mentions & \ref{fig:user_replies_and_mentions} & $\bigcirc$ & $\mathcal{B}$* & $\mathcal{B}$* & $\mathcal{B}$ & 99\% \\ \hline
	{\em More} URLs in tweets & \ref{fig:url_content_uploading} & $\mathcal{B}$** & $\mathcal{B}$** & $\mathcal{B}$** & $\mathcal{B}$** & 99\% \\ \hline
	{\em More} CDN content (KByte) uploaded & \ref{fig:content_size} & $\mathcal{B}$** & $\mathcal{B}$** & $\mathcal{B}$** & $\mathcal{B}$** & 95\% \\ \hline
	{\em Higher} likes received per tweet & \ref{fig:likes_per_tweet} & $\mathcal{H}$** & $\mathcal{H}$** & $\mathcal{H}$** & $\mathcal{B}$ & 99\% \\ \hline
	{\em Higher} retweets received per tweet & \ref{fig:retweets_per_tweet} & $\mathcal{H}$** & $\mathcal{H}$** & $\mathcal{H}$** & $\mathcal{B}$ & 99\% \\ \hline
	{\em More} tweets favourited (liked) & \ref{fig:favourited_by_user} & $\mathcal{B}$** & $\mathcal{H}$** & $\mathcal{H}$** & $\mathcal{H}$** & 99\% \\ \hline
	{\em Higher} follower-friend ratio & \ref{fig:follower_friend_ratio} & $\mathcal{B}$** & $\mathcal{B}$* & $\mathcal{B}$* & $\mathcal{B}$** &  \\ \hline
	{\em More} activity sources & \ref{fig:sources_count} & $\mathcal{B}$* & $\mathcal{B}$ & $\mathcal{B}$ & $\mathcal{B}$ & 99\% \\ \hline
	\end{tabular}
	\label{tab:featureinclination}
\end{table}

The {\em t-test} is generally used to determine if two sets of data (bots and humans in this case) for each feature are significantly different from one another.
This confirms a number of noteworthy behavioural differences between bots and humans in this characterisation study.
For example, confirming past work, we observe a significant and longstanding bot presence on Twitter (\S\ref{subsec:accountage}).
We show that bot users in \gonem tweet more often than humans in the same category.
However, there are a small number of features for which humans and bots are more similar, particularly in \gtenm, where we find humans often exhibit bot-like activity (\eg tweeting a lot).

That said, overall, we find that bot accounts generate larger amounts of tweets than their human counterparts, whilst they (proportionally) rely far more heavily on retweeting existing content and redirecting users to external websites via URLs (\S\ref{subsec:contentcreation}).
These trends appear to be driven by their difficulty in creating novel material when compared to humans.
Despite the voluminous amount of tweets contributed by bots, we find that they still fail to beat humans in terms of {\em quality} though, \eg humans receive a mean of 19$\times$ and a median of 41$\times$ more likes per tweet than bots across all popularity groups (\S\ref{subsec:contentpopularity}).
The divergence between retweet rates is even greater.
Humans receive a mean of 10$\times$ and a median of 33$\times$ more retweets per tweet than bots.

Regardless of their reputed sophistication, it seems that bots still fail to impress users when viewed through the lens of these metrics.
Bots also seem to spend less time liking others' tweets (\S\ref{subsec:contentconsumption}); in other words, humans see greater value in showing their appreciation of other content and users.
This trend is mirrored in other metrics too, \eg humans tend to have a higher proportion of reciprocal relationships (friendships) than bots (\S\ref{subsec:accountreciprocity}).
When combined, we believe that these propensities are heavily influenced by a form of social obligation that bots clearly do not feel.
The lack of such considerations may lead to dramatic changes in social structures and interactions in the longterm (as the bot population increases).

It was also particularly interesting to note the similarities between celebrity users and bots; we even found that some celebrities use bot-like tools (\S\ref{subsec:tweetgenerationsources}).
Finally, we found the two user groups exhibit homophily both in terms of one-hop interactions and wider communities.
For example, bots retweeting other bots is over 3$\times$ more regular than bots retweeting humans, whereas humans retweeting other humans is over 2$\times$ greater (\S\ref{subsec:contentpopularity}).


There is a series of interesting and open questions that need to be explored and answered in our future work.
A major line of future work is expanding the size of our data to understand how bots impact information dissemination more widely.
Currently, we only focus on the accounts that actively disseminate content (\ie tweets or retweets).
It would be worthwhile to also inspect less active bots and humans, \eg fewer than 1K followers.
Moreover, we wish to go beyond bot \vs human studies to understand the \emph{purpose} of the (bot) accounts, \eg news disseminators, socio-political infiltrators or activists, marketing bots, spam and malicious content propagators, \etc.
We also noticed a non-negligible amount of bots that create original content \cite{freitas2015reverse}.
Thus, it would be interesting to further explore the content creation process and perform topical analysis, \ie how and what material do bots tweet, and which approaches gain most traction.
It would be interesting to further explore the content creation process and perform topical analysis, \ie how and what material do bots tweet and which approaches gain most traction, especially for bots that create original content.
Finally, our investigations have revealed inaccuracies in existing bot detection algorithms (hence our use of manual classification).
Manual approaches, however, reduce the ability to study very large number of accounts.
Thus, utilising our findings to fine-tune an automated but reliable bot detection tool is a major remaining challenge.
These topics are reserved as our future research, and plan to extend our measurement platform accordingly to help us seek answers.

\bibliographystyle{abbrv}
\begin{small}
\bibliography{main}
\end{small}

\end{document}